\documentclass[a4paper,12pt,oneside,english]{report}
\usepackage{natbib} 
\usepackage{mathptmx}
\usepackage{amsmath}
\usepackage{amsfonts}
\usepackage{amssymb}
\usepackage{rotating}
\usepackage{subfigure}

\usepackage{graphicx}
\usepackage{babel}
\usepackage{epsfig}

\def\mnras{{\it Mon. Not. R. Astron. Soc.}}

\def\aapr{\it Ann. Rev. Astron. Astrophys.}

\makeatother
\begin{document}
\noindent Submitted to {\bf{\textit{J. Astrophys. Astr. - 2011}}}\\ 
\line(1,0){480}
\vspace{0.2cm}\\\\
{\LARGE{\bf{\noindent Deep GMRT 150 MHz observations of the DEEP2 fields: Searching for High Red-shift Radio Galaxies Revisited}}}

\begin{center}
Susanta Kumar Bisoi${^1}$, C.H. Ishwara-Chandra${^2}$, S. K. Sirothia${^2}$, P. Janardhan${^1}$ 
\end{center}
$\;\;$ \\
\noindent ${^1}$ {{Physical Research Laboratory, Ahmedabad 380 009, India. \\email: susanta@prl.res.in  email: jerry@prl.res.in}} \\\\
\noindent ${^2}$ {{National Centre for Radio Astrophysics, TIFR, Post Bag No. 3, Ganeshkind, Pune 411 007, India. email: ishwar@ncra.tifr.res.in  
email:sirothia@ncra.tifr.res.in}} \\\\

\section*{Abstract}
{\it{High red-shift radio galaxies are best searched at low radio frequencies, due to its steep radio 
spectra. Here we present preliminary results from our programme to search for high red-shift 
radio galaxies to $\sim$ 10 to 100 times fainter than the known population till date. We have 
extracted ultra-steep spectrum (USS) samples from deep 150 MHz Giant Meter-wave Radio Telescope 
(GMRT) observations from one of the three  well-studied DEEP2 fields to this affect. From correlating 
these radio sources w.r.t to the high-frequency catalogues such as VLA, FIRST, and NVSS at 1.4 GHz, 
we find $\sim$ 100 steep spectrum (spectral index, $\alpha$ $>$ 1) radio sources, which are good 
candidates for high red-shift radio galaxies.}}

\section*{Introduction}\label{sec:intro} 
The high red-shift radio galaxies (HzRGs) have emerged as a unique probe to cosmic evolution 
containing important information on the physics, and the properties of the early Universe. 
It had been well established that radio sources with steep radio spectra are more distant 
than sources with normal spectra. This correlation between $\alpha$ and \textit{z} has been 
exploited in the last three decades \cite{MDe08} which yielded about 45 radio 
galaxies at z $>$ 3 till date. The median flux density of all of these known HzRGs is 1.3 Jy 
at 150 MHz which is two orders of magnitude above the detection limit of the GMRT at 150 MHz. 
The GMRT 150 MHz has high angular resolution of $\sim$ $20''$, and better sensitivity $\sim$ 
1 mJy, suggesting that a large number of HzRGs, two orders of magnitude less luminous than 
all of the known objects, can be detected with GMRT \cite{IsS10}. To this 
effect, we have carried out deep 150 MHz observations of three of the four DEEP2 (Deep 
Extragalactic Evolutionary Probe 2) fields, which has optical spectra of $\sim$ 50,000 objects.
\section*{Result and Discussion}
\label{sec:res}
One of the deep DEEP2 fields centered on 2330+0000 was analysed using AIPS++ \cite{SaS09}, 
and the final image obtained has an rms noise $\sim$ 1 mJy/beam with a resolution of $20''$. 
The 150 MHz source catalog of this field contains $\sim$ 400 radio sources to $20\%$ peak primary 
beam response with flux density limit of $\sim$ 6 mJy while the median flux density is $\sim$ 100 mJy. 
The spectral index is estimated by comparison of 150 MHz flux density with that of NVSS and FIRST at 
1.4 GHz. A total of 368 sources from GMRT 150 MHz catalog has counterparts found at FIRST and/or NVSS. 
Figure 1 shows the spectral index distribution of sources as a function of 150 MHz flux density. 
Above a cutoff of $\alpha$ = 1, there are about 100 steep spectrum sources. The FIRST position of 
these sources will be used to search SDSS and DEEP2 optical catalogs to eliminate nearby radio sources, 
and the unidentified radio sources will be the potential candidates for HzRGs.
\protect\begin{figure}[ht]
\vspace{16.0cm}
\includegraphics{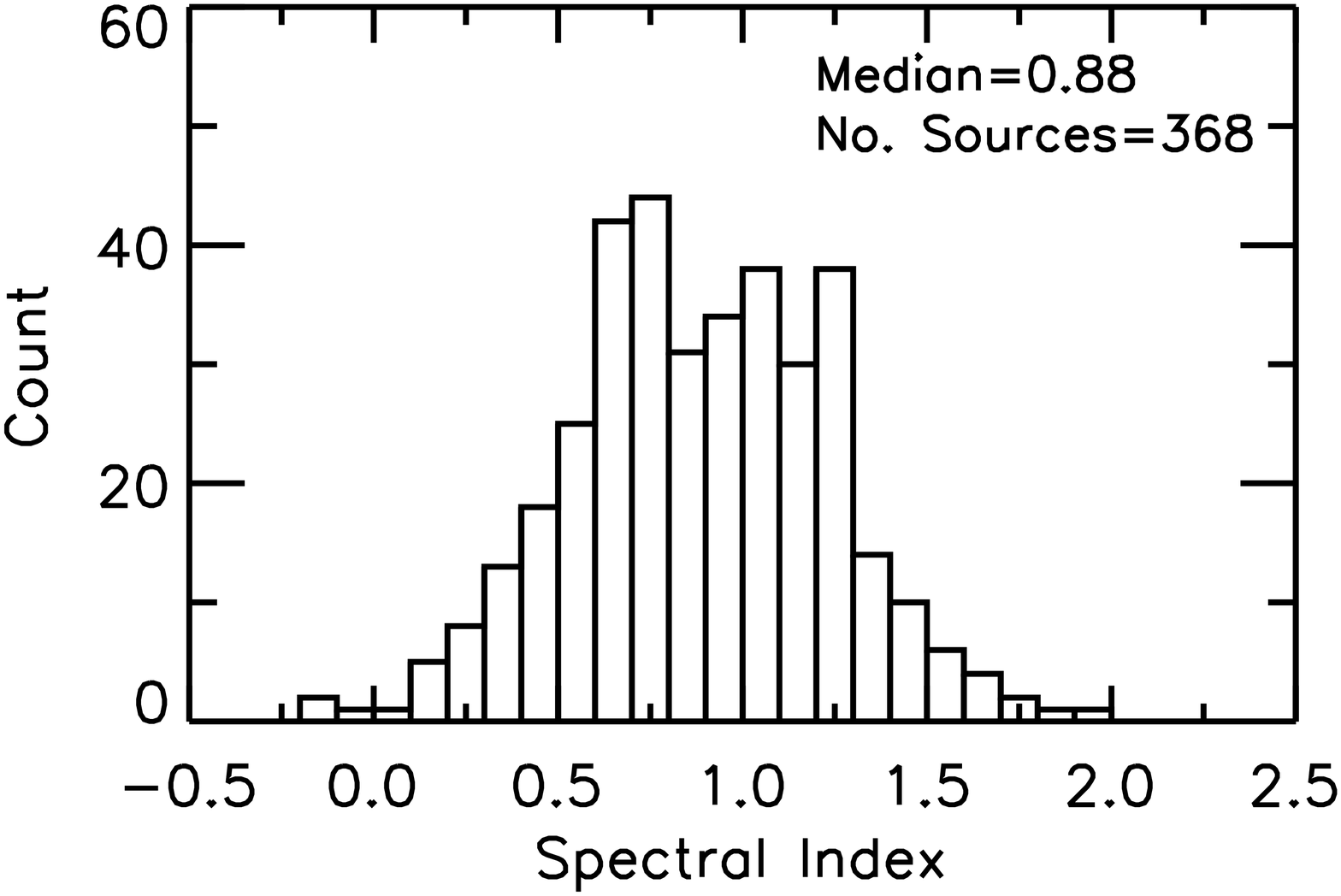}
\includegraphics{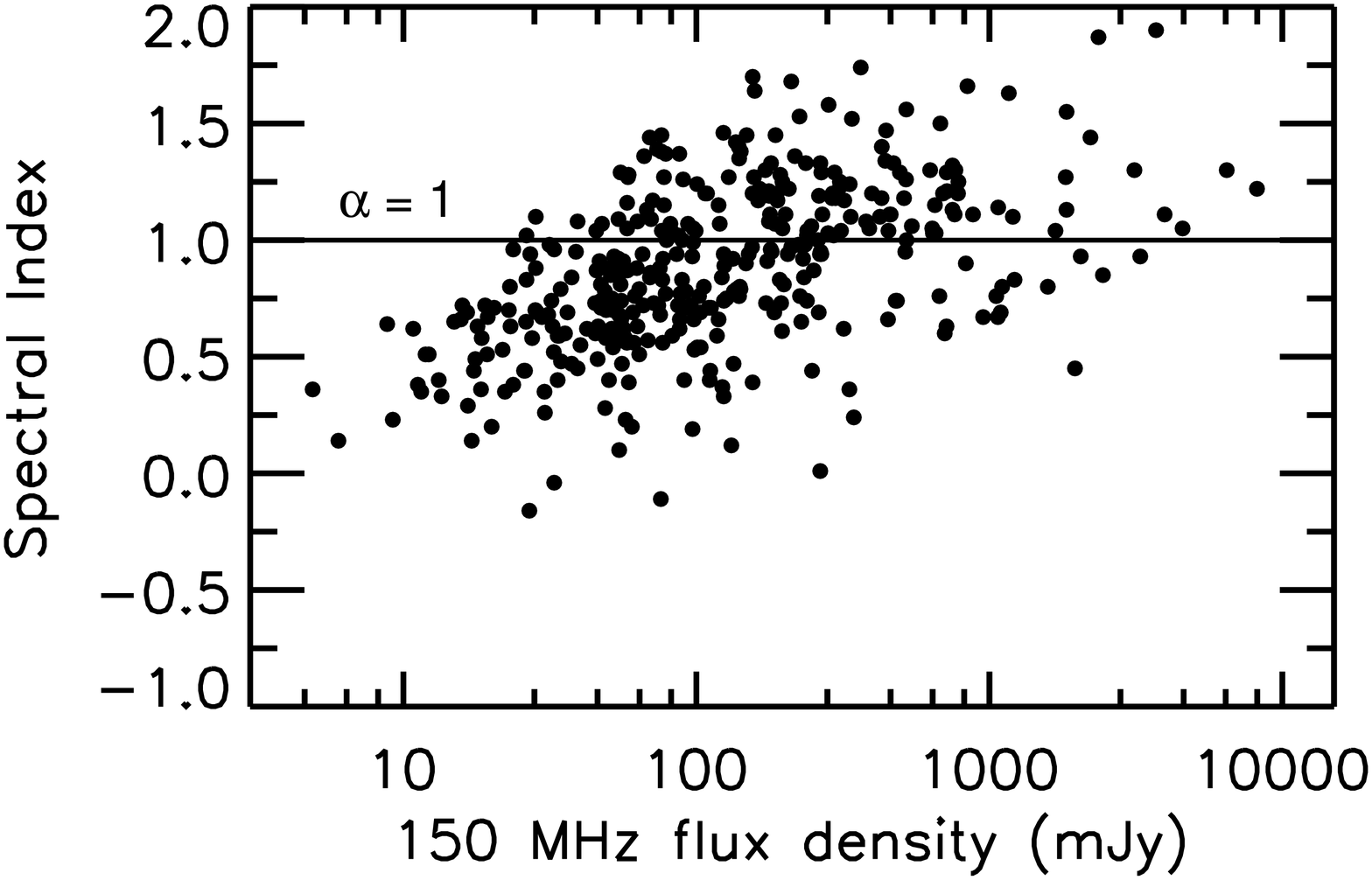}
\caption{Histogram of spectral index of sources 
(Top) and spectral index distribution of sources as a function of flux density at GMRT 150 MHz (Bottom)}
\label{fig:collage2}
\end{figure}	  
%
%
%

\begin{thebibliography}{}

\bibitem[{\it {Ishwara-Chandra} et~al.}(2010)]{IsS10}
{Ishwara-Chandra}, C.~H., S.~K. {Sirothia}, Y.~{Wadadekar}, S.~{Pal}, and
  R.~{Windhorst}, {Deep GMRT 150-MHz observations of the LBDS-Lynx region:
  ultrasteep spectrum radio sources}, {\it \mnras}, {\it 405}, 436--446, 2010.

\bibitem[{\it {Miley} and {De Breuck}}(2008)]{MDe08}
{Miley}, G., and C.~{De Breuck}, {Distant radio galaxies and their
  environments}, {\it \aapr}, {\it 15}, 67--144, 2008.

\bibitem[{\it {Sirothia} et~al.}(2009)]{SaS09}
{Sirothia}, S.~K., D.~J. {Saikia}, C.~H. {Ishwara-Chandra}, and N.~G.
  {Kantharia}, {Deep low-frequency observations with the Giant Metrewave Radio
  Telescope: a search for relic radio emission}, {\it \mnras}, {\it 392},
  1403--1412, 2009.

\end{thebibliography}
%

%
\end{document}